\documentclass[12pt]{iopart}

\newcommand{\mb}[1]{\mbox{\boldmath $#1$}}
\newcommand{\npb}{$\{\mb{k},\mb{\ell},\mb{m},\mb{\bar{m}}\}\;$}
\def \ts {\textstyle}

\begin{document}

\jl{6}

\title[On the Papapetrou field in vacuum]
{On the Papapetrou field in vacuum}

\author{Francesc Fayos\dag$\S$\ and Carlos F. Sopuerta\ddag
\footnote[3]{Also at the Laboratori de F\'{\i}sica Matem\`atica, 
Societat Catalana de F\'{\i}sica, I.E.C., Barcelona, Catalonia, 
Spain. E-mail: labfm@ffn.ub.es\,, cfs@tpi.uni-jena.de}}

\address{\dag\ Departament de F\'{\i}sica Aplicada, UPC, 
E-08028 Barcelona, Spain}

\address{\ddag\ Institut for Theoretical Physics, FSU Jena, 
Max-Wien-Platz 1, D-07743 Jena, Germany}

\begin{abstract}
In this paper we study the electromagnetic fields generated by a 
Killing vector field in vacuum space-times ({\it Papapetrou fields}). 
The motivation of this work is to provide new tools for the resolution of 
Maxwell's equations as well as for the search, characterization, and
study of exact solutions of Einstein's equations. 
The first part of this paper is devoted to an algebraic study in which 
we give an explicit and covariant procedure to construct the principal 
null directions of a Papapetrou field.  In the second part, we
focus on the main differential properties of the principal directions,
studying when they are geodesic, and in that case we compute their
associated optical scalars.  With this information we get the conditions
that a principal direction of the Papapetrou field must satisfy in order
to be aligned with a multiple principal direction of the Weyl
tensor in the case of algebraically special vacuum space-times.
Finally, we illustrate this study using the Kerr, Kasner and 
{\it pp waves} space-times.
\end{abstract}




\section{Introduction}

In a relevant and historic paper, Papapetrou~\cite{PAPA} pointed out 
that a Killing vector field (KVF), say $\mb{\xi}$, can be considered 
as the vector potential generating an electromagnetic field with current
$j^a=R^a{}_b\xi^b$ and satisfying a covariant version of the Lorentz 
gauge.  As an immediate consequence, in vacuum space-times this 
electromagnetic field, which we will call here the {\em Papapetrou field}, 
satisfies the Maxwell equations in the absence of electromagnetic 
sources.

The Papapetrou fields have been used in the search and study of exact 
solutions of Einstein's field equations (see e.g.~\cite{HOMI,CAMI}), 
and in the study of black holes under external electromagnetic 
fields~\cite{WALD}.  In particular, it turns out that the Kerr-Newman 
electromagnetic field is a Papapetrou field generated by the
timelike KVF of the Kerr metric (for more information see 
references~\cite{WALD,MTWB}).

The applications and motivation of this paper concern with issues 
in which the Papapetrou field plays a central role.  
The first one is the resolution of Maxwell's equations in a 
curved space-time, which is a subject of special relevance for
instance in the study of electromagnetic perturbations of black holes 
(an exhaustive account is given in~\cite{CHAN}).  When write Maxwell 
equations in a null basis, using the Newman-Penrose formalism~\cite{NEPE} 
(see also~\cite{CHAN,KSHM}), we get a set of first-order partial 
differential equations for the complex components of the electromagnetic 
field, namely $\Phi_A$ ($A=0,1,2$).  The integrability conditions for 
the complex component $\Phi_1$ (the {\it conditional} system) form in 
general a third-order differential system (with respect to the components
$\Phi_0$ and $\Phi_2$), however, as it was shown in~\cite{CFF1}, if we 
choose the Newman-Penrose basis to be adapted to the principal null 
directions (or null eigendirections) of any particular regular solution 
of Maxwell's equations, the resulting conditional system is of second order.  
This is an important fact in order to integrate Maxwell's equations.
For instance, as it was remarked in~\cite{CFF1}, this is just what 
happens with the Teukolsky-Press relations~\cite{TEUK,TEPR}, which
were the starting point for the integration of Maxwell's equations
in perturbed Kerr space-times (see~\cite{CHAN} and references therein).
In~\cite{CFF1}, it is shown that the Teukolsky-Press relations, 
completed with an additional equation, constitute a conditional 
system of second order for Maxwell's equations (in the variables $\Phi_0$ 
and $\Phi_2$).  This is due to the fact that they were
written using a Newman-Penrose basis adapted to the principal directions
of the Kerr space-time, which are also the principal directions of
a regular electromagnetic field (see also~\cite{CFF1}).
These facts emphasize the importance of knowing a particular regular 
solution of Maxwell's equations.  In general space-times we do not have 
a method to construct such a solution, but in the case of vacuum 
space-times possessing a KVF, a particular solution is given by the
Papapetrou field.  Hence, part of this paper is devoted to make an
exhaustive algebraic study of these electromagnetic fields, giving the 
procedure to construct a Newman-Penrose basis in which the integrability
conditions of the Maxwell equations are directly second-order partial 
differential equations.  For the sake of completeness we will also
treat the case of singular Papapetrou fields,  which appear 
in relevant space-times like the well-known {\em pp waves}.

On the other hand, Papapetrou fields provide us a link between the
Killing symmetries and the algebraic structure of the space-time,
which is a subject treated only occasionally in the literature 
(see~\cite{KSHM}, Chapter 33).   This relationship can be found
by studying the alignment of the principal directions of the 
Papapetrou field with those of the gravitational field, i.e., of
the Riemann tensor (which in vacuum reduces to the Weyl tensor).
In this sense, a remarkable example is the Kerr metric~\cite{KERR}.
In~\cite{FASO,MARS} it is shown that the principal directions of the 
Papapetrou field associated with the timelike KVF coincide with the two 
repeated principal directions of the space-time (it is Petrov type D).   
This link between symmetries and algebraic structure provides a powerful
tool for the study and search of vacuum gravitational fields.  
As an example, a new characterization of the Kerr metric can be
found.  In a recent work by Mars~\cite{MARS}, it is shown
that the Kerr metric is the only stationary asymptotically-flat vacuum 
space-time in which the aforementioned alignment is given.
This characterization of the Kerr metric is 
obtained everywhere, not only where the Killing is timelike. 
Here, we will see an alternative and simple way of characterizing 
the Kerr metric by using our formalism and the connection with the 
eigenray formalism of Perj\'es~\cite{PER2}.

These facts motivate the study of the differential properties
of the principal directions of the Papapetrou field.  In particular,
the study of under which conditions they are geodesic and shear-free 
will tell us, via the well-known Goldberg-Sachs theorem~\cite{GOSA},  
when a principal direction of the Papapetrou field is aligned with a
multiple principal direction of an algebraically special vacuum
gravitational field.   As we will see in the development of the paper,
these properties, and other properties related with the optical
scalars of the principal directions, can be determined only in terms 
of the principal direction itself and the Ernst potential associated 
with the KVF under consideration.   This result is relevant in the sense
that it provides a way of making interesting Ans\"atze for the
search of new solutions of Einstein's equations which, in the light
of the example of the Kerr metric and other examples that will be 
given in this paper, can be physically well motivated.

The plan of this paper is the following:  in Section~\ref{dosf}, we 
give some general properties of antisymmetric tensors (2-forms), which 
will be needed for the development of the next sections. In 
Section~\ref{sepf}, we introduce the Papapetrou field associated with a
KVF. Then, we make an exhaustive study of its algebraic structure,
for which we shall distinguish between regular and singular fields. 
In the regular case, we will introduce a complex 1-form, proportional
to the differential of the Ernst potential, which will play
a central role in our study. Then, we will determine explicitly
(in terms of quantities defined only from the KVF and the metric)
the eigenvalues, the principal null directions, and the orthogonal 
2-planes to these directions (the Maxwellian structure).  In the 
singular case we determine the only principal null direction.
In Section~\ref{prop}, we use this algebraic study to analyze the 
main differential properties of the principal direction(s). 
We will give the conditions for a principal direction to be 
geodesic, and for that case we will give the expression for the
optical scalars, which can be written in terms of the complex
1-form mentioned above and the principal direction.
An illustration of how this study works is given in Section~\ref{exam}, 
where some examples are discussed, in particular the case of the Kerr 
metric~\cite{KERR} is treated with great detail.  
In Section~\ref{reco}, we discuss the main results and consequences 
of this work, as well as possible extensions.  Finally, in~\ref{appa} we 
summarize the known results about the case of a null Killing vector,  
and in~\ref{appb} we give the connection with the {\em eigenray} 
formalism of Perj\'es~\cite{PER2}. Through this paper we will follow 
the notation and conventions of~\cite{KSHM} unless otherwise stated.

\section{Algebraic structure of a 2-form\label{dosf}}

In this section we review briefly the algebraic structure of
a 2-form $\mb{F}$ (an antisymmetric tensor),  describing
an explicit and covariant procedure to construct the null principal 
direction(s) (eigendirections).  First, we deal with the regular 
(non-singular) case, that is, when at least one of the two invariants 
is not zero:
\[\tilde{F}^{ab}\tilde{F}_{ab} \neq 0 \,, \hspace{1cm} 
\tilde{F}_{ab} \equiv F_{ab}+ i *F_{ab}\,, \hspace{1cm}
*F_{ab} \equiv \frac{1}{2}\eta_{abcd}F^{cd} \,, \]
where $*$ and $\,\tilde{}\,$ denote the dual and self-dual operations 
respectively, and $\eta_{abcd}$ are the components of the volume
4-form. At the end of this section we will extend the procedure to 
the singular case.

A regular 2-form $\mb{F}$ can be completely characterized in terms of 
its eigenvalues $(\alpha,\beta)$ and null principal directions 
$(\mb{k},\mb{\ell})$. In fact, it can be written as follows
\begin{equation}
\mb{F} = \alpha\, \mb{G} -\beta\, * \mb{G}\,, \hspace{5mm}
\mb{G}\equiv \mb{k}\wedge\mb{\ell}\,, \hspace{5mm} k^a\ell_a =-1 \,.
\label{2for}
\end{equation}
Note that $\mb{G}$ is a 2-form of rank 2 [$\det(\mb{G})=0$].  
Moreover, the corresponding dual and self-dual 2-forms $*\mb{F}$ 
and $\mb{\tilde{F}}$ are given by
\begin{equation}
*\mb{F} = \beta\, \mb{G}+\alpha\, * \mb{G}\,, \hspace{8mm}
\mb{\tilde{F}}=(\alpha+i\beta)\mb{\tilde{G}} \,, \label{d2fo} 
\end{equation}
from where we find the following relation between the invariants
of $\mb{F}$ and its eigenvalues
\begin{equation} 
\tilde{F}^{ab}\tilde{F}_{ab} = -4 (\alpha+i\beta)^2 \,. \label{reig}
\end{equation}
Given a 2-form $\mb{F}$, this relation allows us to find the
eigenvalues ($\alpha$,$\beta$), and from (\ref{d2fo}) we can find
the singular 2-form $\mb{G}$.
On the other hand, for each 2-form $\mb{F}$ we can construct the 
following symmetric tensor
\begin{equation} 
T_{ab} = \frac{1}{2}\left(F_a{}^cF_{bc}+*F_a{}^c*F_{bc}\right) 
\,. \label{emte} 
\end{equation}
When $\mb{F}$ represents an electromagnetic field, this tensor is
the energy-momentum tensor.  In the case of a regular 2-form, using 
(\ref{2for},\ref{d2fo}) we find an alternative expression for 
$T_{ab}$ 
\begin{equation}
T_{ab} = (\alpha^2+\beta^2)\left[k_{(a}\ell_{b)}+
m_{(a}\bar{m}_{b)}\right] \, , \label{emt2} 
\end{equation}
where $\mb{m}$ is a complex vector orthogonal to the 2-planes
generated by the principal directions $\mb{k}$ and $\mb{\ell}$, 
and such that \npb is a Newman-Penrose basis. Therefore, this basis 
is adapted to the Maxwellian structure of the 2-form $\mb{F}$.  We 
can write the dual of $\mb{G}$ in terms of $\mb{m}$ and its
complex conjugate $\mb{\bar{m}}$ as 
$*\mb{G}=i\mb{m}\wedge\mb{\bar{m}}$.
Note that expression (\ref{emt2}) shows the 2+2 decomposition of the 
energy-momentum tensor (equivalently of the Ricci tensor) of a 
regular 2-form, or in other words, that the Segr\'e type is 
[(1 1) (1\,,\,1)] (see for instance~\cite{KSHM}).

The problem of determining the principal directions of a 2-form
has been treated by Coll and Ferrando~\cite{COFE}.  To that end they
introduced the so-called {\it concomitant} of a 2-form 
\begin{equation}
{\cal F}_{ab} \equiv \alpha F_{ab} +\beta *F_{ab} - T_{ab} +
\frac{1}{2}(\alpha^2+\beta^2)g_{ab} \,. \label{fgte}
\end{equation}
Using this tensor, the principal null directions are given by
\begin{equation} 
K_a = {\cal F}_{ab}U^b \,, \hspace{1.5cm}
L_a = {\cal F}_{ba}U^b \,,  \label{pdfo}
\end{equation}
where $\mb{U}$ is an arbitrary timelike vector field.  We can 
understand this property by rewriting the tensor ${\cal F}$
in the following form
\begin{equation} 
{\cal F}_{ab} = (\alpha^2+\beta^2)\left(G_{ab}+G_a{}^cG_{cb}\right) 
= -2(\alpha^2+\beta^2)\ell_ak_b \,, \label{tefg}
\end{equation}
where we have used (\ref{2for}-\ref{emte}). As we can see from this 
expression, $\mb{K}$ and $\mb{L}$ are parallel to the null vector 
fields $\mb{k}$ and $\mb{\ell}$ respectively [see expressions 
(\ref{2for})].   It is important to point out that this 
method provides a covariant way of finding the principal 
directions of a 2-form. 

On the other hand, in order to determine the 2-planes orthogonal to
the principal directions, we can introduce another tensor 
\begin{equation}
{\cal F}^\bot_{ab}\equiv i(\beta F_{ab}-\alpha *F_{ab})+
T_{ab}+\frac{1}{2}(\alpha^2+\beta^2)g_{ab}\,. \label{fort}
\end{equation}
From this complex tensor we can find a complex vector
$\vec{\mb{m}}$ describing the orthogonal 2-planes.  To that
end, we have to contract (\ref{fort}) with any spacelike vector 
field linearly independent from $\mb{k}$ and $\mb{\ell}$, and
normalize the result ($m^a\bar{m}_a=1$).  We can show this fact
by rewriting ${\cal F}^\bot_{ab}$ in terms of the Newman-Penrose 
basis \npb
\[ {\cal F}^\bot_{ab} = -(\alpha^2+\beta^2)(i*G_{ab}+
*G_a{}^c*G_{cb})=-2(\alpha^2+\beta^2)m_a\bar{m}_b\,, \]
Note that $\mb{m}$ is fixed up to a factor $\mbox{e}^{iC}$, 
where $C$ is an arbitrary real scalar. 

To sum up, we can construct the Maxwellian structure associated 
with a regular 2-form from tensors (\ref{fgte}) and (\ref{fort}).

In the singular case, the self-dual 2-form $\tilde{\mb{F}}$ can 
always be written as follows (see~\cite{KSHM})
\[\tilde{F}_{ab}=4\Phi k_{[a}m_{b]}\,,\hspace{1cm}k^a m_a=0\,,\] 
where $\mb{m}$ is some normalized null complex vector field ($m^am_a=0$
and $m^a\bar{m}_a=1$) and $\Phi$ is a complex scalar field.
As is well-known, $\mb{k}$ is a multiple eigenvector with zero
eigenvalue. In addition, the energy-momentum tensor (\ref{emte})
now takes the following form
\[ T_{ab} = 2\Phi\bar{\Phi}k_ak_b \,, \]
that is, the Segr\'e type is [(1 1\,,\,2)].  We can use the 
previous procedure to find explicitly the principal
direction only by considering the fact that the eigenvalues are 
identically zero ($\alpha=\beta=0$). Therefore, the tensor 
${\cal F}_{ab}$ is simply $-T_{ab}$, showing that there is only one 
principal direction ($\mb{K}$ and $\mb{L}$ coincide), which is 
determined by
\[ K_a = T_{ab}U^b \,,\]
where $U^a$ is an arbitrary timelike vector field.  In that case,
the tensor ${\cal F}^\bot_{ab}$ does not provide any information 
since there is not a 2+2 structure, which is clear from
the Segr\'e type of the energy-momentum tensor.

\section{The Papapetrou field\label{sepf}} 
 
Let $(V_4,g)$ be an arbitrary vacuum ($R_{ab}=0$) space-time 
(see~\cite{HAEL} for details) endowed
with a one-dimensional isometry group generated by a non-null KVF
$\mb{\xi}$
\begin{equation}
\xi_{a;b}+\xi_{b;a}=0 \,. \label{kill} 
\end{equation}  
That is, we will assume that $\mb{\xi}$ has a fixed character 
(timelike or spacelike) in an open domain of the space-time\footnote{The
case of a null KVF is of little interest since all the vacuum 
space-times admitting a null KVF are known, they can be found
in~\cite{KSHM} (section 21.4; see also Ref.~\cite{PER1} 
and~\ref{appa}).}. In this domain, the electromagnetic field, 
{\em Papapetrou field},  generated by $\mb{\xi}$ is given by~\cite{PAPA} 
\begin{equation} 
F_{ab} = (\mb{d\xi})_{ab}=\xi_{b;a}-\xi_{a;b} = 2\xi_{b;a} 
\,, \label{papa} 
\end{equation} 
where we have used the Killing equations (\ref{kill}) and $\mb{d}$ 
denotes exterior differentiation. Note that $\mb{F}$ is
defined up to a multiplicative constant.  Moreover, taking into account
the following relation for the second covariant derivatives of a KVF
\begin{equation} 
\xi_{a;bc} = R_{abcd}\xi^d \,, \label{ridt}
\end{equation}
which comes from the Ricci identities, we can show that this 
2-form $F_{ab}$ satisfies the Maxwell
equations in the absence of electromagnetic charge and current
distributions
\begin{equation} 
F_{[ab;c]}=0, \quad F^{ab}{}_{;b}=0 \,. \label{maxe} 
\end{equation}
In addition, from the definition of the Papapetrou field (\ref{papa}),
we can see that the KVF $\mb{\xi}$ plays the role of the
electromagnetic vector potential, and from the Killing equations
(\ref{kill}), that it satisfies the covariant Lorentz condition
$\xi^a{}_{;a}=0$.

Some important quantities associated with a KVF are:
the {\em norm} 
\[ N\equiv\xi^a\xi_a \,, \]
which we have assumed to be different from zero, the {\em orthogonal 
projector}
\[ h_{ab}\equiv g_{ab}-\frac{1}{N}\xi_a\xi_b\,, \hspace{1cm}
h_{ab}\xi^b = 0 \,, \] 
the gradient of the norm 
\[\psi_a\equiv-(\mb{dN})_a=-N_{,a}\,, \hspace{1cm} 
\psi^a\xi_a=0 \,,  \] 
and the {\em twist} 
\[\omega_a\equiv-*(\mb{\xi}\wedge\mb{d\xi})_a=\eta_{abcd}\xi^b
\xi^{c;d}\,, \hspace{1cm} \omega^a\xi_a=0 \,.  \]
The twist has three independent 
components which are also contained in the following 2-form
\begin{equation} 
W_{ab}=h_a{}^ch_b{}^d\xi_{[c;d]}=h_a{}^ch_b{}^d\xi_{c;d} \,, 
\hspace{1cm} W_{ab}\xi^b= 0 \,, \label{w2fo}
\end{equation}  
usually called the {\em rotation} of $\mb{\xi}$.  The rotation
is related to the twist through the expressions
\[ \omega_a = \eta_{abcd}\xi^b W^{cd}\, , \hspace{2cm}
W_{ab} = \frac{1}{2N}\eta_{abcd}\xi^c \omega^d \, . \]

In terms of these quantities, the Papapetrou field can be written in
the following way
\begin{equation}
F_{ab}= \frac{2}{N}\left(\xi_{[a}\psi_{b]}-N W_{ab}\right) \,,
\hspace{6mm} *F_{ab}= \frac{2}{N}\left(\xi_{[a}\omega_{b]}+N 
M_{ab}\right) \,, \label{papf}
\end{equation}
where the definition of $M_{ab}$ is
\[ M_{ab} \equiv \frac{1}{2N}\eta_{abcd}\xi^c \psi^d \,. \]

On the other hand, it is well-known that in vacuum space-times the twist 
1-form satisfies $\mb{d\omega}=\mb{0}$ (see~\cite{KSHM} for instance), 
and therefore there is locally a function $\Omega$ so that $\mb{\omega} 
= \mb{d\Omega}$. 
Then, we can introduce the Ernst potential (see~\cite{ERNS,KSHM})
associated with $\mb{\xi}$ 
\[ {\cal E} = - N + i\Omega \,. \]
In terms of this potential, the self-dual Papapetrou field can be cast
into the following form
\begin{equation}
\mb{\tilde{F}} = \frac{1}{N}\left(\mb{\xi}\wedge\mb{\epsilon}
\right)\mb{\tilde{}}\,,\hspace{10mm}\mb{\epsilon}\equiv\mb{d{\cal E}}=
\mb{\psi} + i \mb{\omega} \hspace{4mm} (\mb{d\epsilon}=0)\,, \label{papk}
\end{equation}
where we have defined $\mb{\epsilon}$ as the exterior derivative of the 
Ernst potential ${\cal E}$.  In vacuum, this 1-form satisfies the 
following equation
\begin{equation} 
\epsilon^a{}_{;a}+N^{-1}\epsilon^a\epsilon_a=0\,, \label{eqep}
\end{equation}
which comes from Einstein's equations. Moreover, from~(\ref{papk}) we
realize that in the case of a timelike KVF, $\mb{\psi}$ and
$\mb{\omega}$ are respectively proportional to the electric and 
magnetic fields measured by observers with 4-velocity 
$\mb{u}=(-N)^{-1/2}\mb{\xi}$.

We begin the process of construction of the principal directions
of the Papapetrou field by the calculation of the eigenvalues
$\alpha$ and $\beta$. Using (\ref{papf}) or (\ref{papk}) we get
\[ \tilde{F}^{ab}\tilde{F}_{ab} = \frac{4}{N}\epsilon^a\epsilon_a = 
\frac{4}{N}\left\{ \psi^a\psi_a - 
\omega^a\omega_a + 2 i \psi^a\omega_a \right\} \,. \]
And comparing this expression with (\ref{reig}) we find
\begin{equation}
\alpha = \frac{s}{\sqrt{2}} \sqrt{ -x + \sqrt{x^2+ y^2} } \,,
\hspace{1.5cm} \beta = -\frac{s}{\sqrt{2}} \sqrt{x + \sqrt{x^2+ y^2} } 
\,,  \label{albe}
\end{equation}
where $x$ and $y$ 
\begin{equation} 
x \equiv \frac{\psi^a\psi_a - \omega^a\omega_a}{N}
\,, \hspace{8mm}  y \equiv \frac{2\psi^a\omega_a}{N} \,, \label{dexy}
\end{equation}
are the two invariants of the Papapetrou
field $\mb{F}$, and $s$ is an arbitrary sign ($s^2=1$).  This sign
reflects the invariance of $\mb{F}$ (\ref{2for}) under the change
$(\mb{k},\mb{\ell},\alpha,\beta)$ $\rightarrow$ 
$(\mb{\ell},\mb{k},-\alpha,-\beta)$.

From (\ref{albe}), we find some interesting relationships 
between the eigenvalues and the invariants
\begin{equation} 
\alpha^2-\beta^2 = -x \,, \hspace{1cm} 2\alpha\beta = -y \,, 
\hspace{1cm} \alpha^2+\beta^2 = \sqrt{x^2+ y^2}\,. \label{reab}
\end{equation}
These relations together with (\ref{dexy}) imply
\begin{equation}
\epsilon^a\epsilon_a = N (x+i y) = -N (\alpha+i\beta)^2 
\,. \label{epsq}
\end{equation}

On the other hand, from (\ref{emte}) we have the following expression
from the energy-momentum tensor of the Papapetrou field (it is important
to remark that this electromagnetic field is a test electromagnetic
field)
\begin{eqnarray} 
T_{ab} = \frac{1}{N^2}\left\{ (\psi^c\psi_c+\omega^c\omega_c)\xi_a
\xi_b + N(\psi_a\psi_b+\omega_a\omega_b)-2N \xi_{(a}\phi_{b)}- 
\right. \nonumber \\ \left. 
\frac{1}{2}N(\psi^c\psi_c+\omega^c\omega_c)g_{ab} \right\} \, , 
\end{eqnarray} 
where
\[ \phi_a \equiv 2 W_{ab}\psi^b = -2M_{ab}\omega^b = 
-\frac{1}{N}\eta_{abcd}\xi^b\psi^c \omega^d  \, . \]

In this situation, we are ready to construct the tensors 
${\cal F}_{ab}$ and ${\cal F}^\bot_{ab}$ (\ref{fgte},\ref{fort}),
and through (\ref{pdfo}) we can find explicit expressions for 
null vector fields tangent to the principal directions of the 
Papapetrou field.  However, in order to carry out these 
calculations, it is necessary to consider separate cases.  First of 
all, we make a division into two groups depending on whether the 
Papapetrou field is regular ($\alpha+i\beta\neq 0$) or singular 
($\alpha+i\beta=0$).  In the first case, we will determine completely 
the characteristic 2+2 structure of a regular 2-form, 
the Maxwellian structure.  In the second case we will give the only 
one principal null direction.

\subsection{Regular Papapetrou fields} \label{repf}
We begin the study of regular Papapetrou fields by introducing a
complex 1-form which will be of crucial importance for the purposes of
this work.  The definition of this 1-form is
\begin{equation}
\mb{E} \equiv \frac{\mb{\epsilon}}{\alpha+i\beta} =
\frac{1}{\alpha+i\beta}(\mb{\psi}+i\mb{\omega}), \hspace{2cm}
\mb{E}\wedge\mb{dE} = 0 \, . \label{defe}
\end{equation}
As we can see, it is proportional to the exterior
derivative of the Ernst potential and hence, it is integrable.  
Moreover, from~(\ref{papk}) we find the following useful relationship 
between $\mb{E}$ and the self-dual 2-form $\mb{\tilde{G}}$ 
[see Eq.~(\ref{d2fo})]
\[ \mb{\tilde{G}} = \frac{1}{N}\left(\mb{\xi}\wedge\mb{E}
\right)\mb{\tilde{}}\hspace{5mm} \Longleftrightarrow \hspace{5mm} 
E_a = -\tilde{G}_{ab} \xi^b \,. \]
On the other hand, $\mb{E}$ satisfies the following equation
\[ E^a{}_{;a}+\frac{E^a(\alpha+i\beta)_{,a}}{\alpha+i\beta} =
\alpha+i\beta \,,\]
which follows from equation (\ref{eqep}).
On the other hand, when it is an exact 1-form, $\mb{dE}=0$, either 
$\alpha+i\beta$ is constant or it is a function only of
the Ernst potential.

The importance of this 1-form is twofold, first it will allow 
us to classify the different subcases that need a separate treatment,
and second, it will serve to write our results in a compact form.
Now, let us split $\mb{E}$ into its real and imaginary parts
\begin{equation}
\mb{E} = \mb{E_R}+i\mb{E_I} \, . \label{reim}
\end{equation}
From the definition (\ref{defe}) we can find the relations between
$(\mb{E_R},\mb{E_I})$ and $(\mb{\psi},\mb{\omega})$
\begin{equation} 
\mb{E_R} = \frac{1}{\alpha^2+\beta^2}(\alpha\mb{\psi}+\beta\mb{\omega})
\, , \hspace{1cm} 
\mb{E_I} = \frac{1}{\alpha^2+\beta^2}(-\beta\mb{\psi}+\alpha\mb{\omega})
\, ,  \label{raip}
\end{equation}
and the converse relations
\[ \mb{\psi} = \alpha\mb{E_R}-\beta\mb{E_I} \, , \hspace{1cm}
\mb{\omega} = \beta\mb{E_R}+\alpha\mb{E_I} \, . \]
From (\ref{defe},\ref{reim},\ref{epsq}) we find the following 
important relation
\begin{equation}
E^aE_a = E_R^a E^{}_{Ra}- E_I^a E^{}_{Ia} +2 i E_R^a E^{}_{Ia} =
- N \,, \label{enor}
\end{equation}
from where we deduce that $\mb{E_R}$ and $\mb{E_I}$ are orthogonal
\begin{equation}
E_R^a E^{}_{Ia} = 0 \,, \label{orto}
\end{equation}
and also, a relation between their norms and the Killing norm (note
that $\mb{E}$ has a real norm)
\begin{equation} 
E_R^a E^{}_{Ra}-E_I^a E^{}_{Ia} = -N \,. \label{keye}
\end{equation}

From these properties we can derive some consequences 
concerning the character of the vector fields
$\{\mb{\xi},\mb{E_R},\mb{E_I},\mb{\phi}\}$. First, when 
$\mb{\xi}$ is timelike ($N<0$), is obvious that $\mb{E_R}$ and 
$\mb{E_I}$ are spacelike, and taking into account that
\begin{equation}
\fl \phi^a\phi_a = \frac{1}{N}\left[(\psi^a\omega_a)^2-(\psi^a\psi_a)
(\omega^b\omega_b)\right] = -\frac{1}{N}(\alpha^2+\beta^2)^2
(E_R^a E^{}_{Ra})(E_I^b E^{}_{Ib}) \,, \label{phsq}
\end{equation}
it is clear that $\mb{\phi}$ is also spacelike.  On the contrary, 
when $\mb{\xi}$ is spacelike ($N>0$), from (\ref{keye}) we have
$E_I^a E^{}_{Ia} = N + E_R^a E^{}_{Ra}$ and then, $\mb{E_I}$ cannot 
be timelike, otherwise $\mb{E_R}$ would be also timelike, in 
contradiction with (\ref{orto}).  Moreover, $\mb{E_I}$ cannot be a 
null vector either, otherwise taking into account (\ref{keye}), which 
implies that $E_R^a E^{}_{Ra}=-N<0$, and (\ref{orto}), $\mb{E_R}$ 
would be also a null vector and using again (\ref{keye}), $\mb{\xi}$ 
should be also a null vector, which contradicts our initial 
assumptions. Therefore, we have shown that {\em $\mb{E_I}$ must be 
always spacelike} (unless it vanishes). In contrast, $\mb{E_R}$ and 
$\mb{\phi}$ can be timelike, or spacelike, or null.  Furthermore, 
when $\mb{E_R}$ is a null vector, $\mb{\xi}$ is spacelike and 
$\mb{\phi}$ is also a null vector parallel to $\mb{E_R}$.

Now, let us introduce some vectors which are adapted to 2+2
decomposition of the energy-momentum tensor of an electromagnetic
field.  These vectors are $\mb{E_R}$, $\mb{E_I}$ and
\begin{equation}
\mb{P}\equiv\frac{1}{N}\left[(E_R^a E^{}_{Ra})\mb{\xi}+\mb{D}\right]
\,, \hspace{1cm} \mb{Q}\equiv\frac{1}{N}\left[(E_I^a E^{}_{Ia})\mb{\xi}+
\mb{D}\right] \, , \label{deeq}
\end{equation}
where $\mb{D}$ is a 1-form defined by 
$\mb{D} \equiv *(\mb{\xi}\wedge\mb{E_R}\wedge\mb{E_I})$. In local
coordinates the components of $\mb{D}$ are 
\[ D_a = \eta_{abcd}\xi^b E_R^c E_I^d = \frac{1}{\alpha^2+\beta^2}
\eta_{abcd}\xi^b\psi^c\omega^d = -\frac{N}{\alpha^2+\beta^2}\phi_a \,.\]
The scalar products between these 1-forms are given by
\begin{equation} 
P^a P_a = - E_R^a E^{}_{Ra} \,,\hspace{5mm} 
Q^a Q_a = E_I^a E^{}_{Ia} \,, \label{ort1}
\end{equation}
\[ P^a E_{Ra} = P^a E_{Ia} = P^a Q_a = Q^a E_{Ra} = Q^a E_{Ia} = 
0 \,, \]
and hence, they form in general an orthogonal basis 
(unless $E_R^a E^{}_{Ra}=0$ or $\mb{E_I}=0$). The vectors $\mb{P}$ 
and $\mb{E_R}$ lie in the 2-planes generated by the principal 
directions.  This fact can be seen from the following properties of 
${\cal F}^\bot_{ab}$
\begin{equation} 
{\cal F}^\bot_{ab}P^b = {\cal F}^\bot_{ab}E_R^b = 0\,,\hspace{10mm}
{\cal F}^\bot_{ab}P^a = {\cal F}^\bot_{ab}E_R^a = 0 \,. \label{forc}
\end{equation}
On the other hand, $\mb{Q}$ and $\mb{E_I}$ determine
in general (when $\mb{E_I}\neq 0$) the orthogonal 2-planes to the 
principal directions. This fact follows from the following relations
\[ {\cal F}_{ab}Q^b = {\cal F}_{ab} E_I^b = 0 \,, \hspace{10mm}
{\cal F}_{ab}Q^a = {\cal F}_{ab} E_I^a = 0 \,. \]
Moreover, since $\mb{E_I}$ is always spacelike and taking into 
account (\ref{ort1}), we deduce that both $\mb{E_I}$ and $\mb{Q}$ are 
spacelike.

In what follows we find explicit expressions for the
principal directions of a regular Papapetrou field and  
determine the 2-planes orthogonal to these directions.  Following  
the previous discussion we will distinguish between 
two subcases, depending on whether or not the norm of $\mb{E_R}$ 
vanishes.

\subsubsection{Generic case $E_R^aE^{}_{Ra}\neq 0$:\label{sbgc}} 

This case represents the generic situation. From (\ref{ort1})
we deduce that either $\mb{P}$ or $\mb{E_R}$ must be timelike and 
hence, in order to obtain the principal directions we can use
these vectors.  However, in this case we can write ${\cal F}$ in 
terms of the basis $\{\mb{P}, \mb{E_R}, \mb{Q}, \mb{E_I}\}$
\[ {\cal F}_{ab} = \frac{\alpha^2+\beta^2}{E_R^cE^{}_{Rc}}
(E_{Ra}+P_a)(E_{Rb}-P_b) \,, \]
from where it follows that the principal directions of the 
Papapetrou field are determined by the following null vector
fields
\begin{equation}
\mb{K} = \mb{P}-\mb{E_R} \,, \hspace{10mm}
\mb{L} = \mb{P}+\mb{E_R} \,. \label{pdlp}
\end{equation}
These expressions also show the fact that in
general $\mb{P}$ and $\mb{E_R}$ determine the 2-planes generated by the
principal directions, as it has been pointed out before [see 
expressions (\ref{forc})].  In order to determine the orthogonal 
2-planes we have to use the tensor (\ref{fort}), which in this 
generic case can be cast into the following form
\[ {\cal F}^\bot_{ab} = \frac{\alpha^2+\beta^2}{E_I^cE^{}_{Ic}}
(E_{Ia}-iQ_a)(E_{Ib}+iQ_b) \,. \]
Then, the orthogonal 2-planes are determined by $\mb{Q}$ and 
$\mb{E_I}$.  Moreover,
as is clear, this expression requires $E^a_IE^{}_{Ia}\neq0$,
equivalent to $\mb{E_I}\neq 0$. In fact, the case $\mb{E_I}=0$ 
needs a separate study, which is given below.

Finally, we want to stress the fact that the 2+2 structure is also 
manifested in the form of the energy-momentum tensor.  In our 
case, we can see this feature by writing $T_{ab}$ using the basis 
$\{\mb{P}, \mb{E_R}, \mb{Q}, \mb{E_I}\}$
\[ \fl T_{ab} = \frac{1}{2}(\alpha^2+\beta^2)
\left\{\frac{1}{E_R^c E^{}_{Rc}}
\left(P_aP_b-E_{Ra}E_{Rb}\right)+\frac{1}{E_I^c E^{}_{Ic}}
\left(Q_aQ_b+E_{Ia}E_{Ib}\right)\right\} \,.   \]

\vspace{2mm}

\noindent$\bullet\,${\em Particular case $\mb{E_I}=0$:} Now we have
$-\beta\mb{\psi}+\alpha\mb{\omega} = 0$
and then $\mb{\psi}$ and $\mb{\omega}$ are linearly dependent, hence
$\mb{D}=0$.  Moreover, this implies $\mb{P}=-\mb{\xi}$ and $\mb{E}=
\mb{E_R}$, therefore we can use the previous expressions 
(\ref{pdlp}).  Then, the
principal null directions of the Papapetrou field are given by
\begin{equation} 
\mb{K} = \mb{\xi}+\mb{E_R} \,, \hspace{15mm}
\mb{L} = \mb{\xi}-\mb{E_R} \,. \label{pdiv}
\end{equation}
On the other hand, the relations $E^a_IE^{}_{Ia}=0$ and $E^a_RE^{}_{Ra}=
-N$ [the last one comes from (\ref{keye})] imply
\[ \psi^a\psi_a+\omega^a\omega_a = -N(\alpha^2+\beta^2) \,, \]
where we have used (\ref{raip}).  Combining the last equation with
expressions (\ref{dexy},\ref{reab}) we arrive at the following results
\[ \psi^a\psi_a=-N\alpha^2\,, \hspace{10mm} \omega^a\omega_a =
-N\beta^2 \,. \]
The peculiarity of this subcase is related with the structure of the 
orthogonal 2-planes.  Now, although we can determine
the 2-planes spanned by the principal null directions, we cannot 
determine explicitly the orthogonal 2-planes, since in this
case we have $\mb{E_I}=\mb{Q}=0$ [see equations (\ref{deeq})].
However, using the results of section 2 we can construct the orthogonal 
2-planes. We only need to contract ${\cal F}^\bot_{ab}$ (\ref{fort}), 
which is now given by
\begin{equation} 
{\cal F}^\bot_{ab}= -\frac{\alpha^2+\beta^2}{N}\left\{
i\eta_{abcd}P^cE^d_R+P_aP_b-E_{Ra}E_{Rb}-Ng_{ab}\right\} 
\,, \label{tenf} 
\end{equation}
with any spacelike vector field independent of $\mb{\xi}$ and $\mb{E_R}$.
Then, after normalization we find $\mb{m}$, which is defined up to a 
factor $e^{iC}$, where $C$ is an arbitrary real scalar. 
Alternatively, given two vector fields, linearly independent and
orthogonal to $\mb{\xi}$ and $\mb{E_R}$, we can construct from them
the orthogonal 2-planes to the principal directions, and therefore
we can find a vector $\mb{m}$.

\subsubsection{Particular case $E_R^aE^{}_{Ra}= 0$:}

Now, let us see what happens when $\mb{E_R}$ is a null vector
\begin{equation}
E^a_RE^{}_{Ra}=0\hspace{8mm} \Longrightarrow \hspace{8mm}
E^a_IE^{}_{Ia}=N\,.  \label{er20}
\end{equation}
Following the discussion given above on the norm of $\mb{E_R}$ and
$\mb{E_I}$, it turns out that this case can only take place when 
$\mb{\xi}$ is spacelike ($N>0$).  

On the other hand, using (\ref{raip}), equations (\ref{keye}) and 
(\ref{er20}) imply
\[ \psi^a\psi_a+\omega^a\omega_a = N(\alpha^2+\beta^2) \,. \]
Combining this result with (\ref{dexy},\ref{reab}) we get expressions
for the norms of $\mb{\psi}$ and $\mb{\omega}$ 
\[ \psi^a\psi_a=N\beta^2\,, \hspace{10mm} \omega^a\omega_a =
N\alpha^2 \,, \]
then, both are spacelike.  Moreover, from (\ref{dexy},\ref{reab}) we
can also deduce the following result
\[ (\psi^a\psi_a)(\omega^a\omega_a) = (\psi^a\omega_a)^2 \,. \]
This is an important relation because it means that either $\mb{\psi}$
and $\mb{\omega}$ are parallel or one of them is a null vector 
orthogonal to the other one.  The first possibility implies that 
$\mb{E_R}=0$, which will be treated later as an special subcase.  The 
second possibility leads to one of the following
two excluding situations: the first one is given by
\begin{equation} 
\psi^a\psi_a=\psi^a\omega_a=\beta=0\,, \hspace{10mm}
\mb{E_R}=\frac{1}{\alpha}\mb{\psi}\,, \hspace{5mm}
\mb{E_I}=\frac{1}{\alpha}\mb{\omega}\,, \label{ern1} 
\end{equation}
and the second one by
\begin{equation}
\omega^a\omega_a=\psi^a\omega_a=\alpha=0\,, \hspace{10mm}
\mb{E_R}=\frac{1}{\beta}\mb{\omega}\,, \hspace{5mm}
\mb{E_I}=-\frac{1}{\beta}\mb{\psi}\,. \label{ern2}
\end{equation}
In both situations (\ref{ern1},\ref{ern2}), equation (\ref{phsq}) 
implies that $\mb{\phi}$ is a null vector field parallel to $\mb{E_R}$, 
and obviously, the same happens with $\mb{D}$. 

With regard to the principal directions, in this case we cannot determine 
completely the 2-planes spanned by the principal null directions, since 
$\mb{P}$ and $\mb{E_R}$ are not linearly independent 
($\mb{P}=N^{-1}\mb{D}$ is also a null vector parallel to $\mb{E_R}$).  
From (\ref{pdlp}) we can see that $\mb{E_R}$ is one of the principal 
directions, but we cannot determine the other one from these expressions.  
This is due to the fact that we have not a timelike vector available.  

Therefore, it is convenient to follow the procedure explained in 
Sec.~\ref{dosf}.  First, we have to construct the tensor 
${\cal F}_{ab}$ (\ref{fgte}). In this case, we can find 
${\cal F}_{ab}$ through the expression (\ref{tefg}) and taking into
account that now we can write
\begin{equation} 
G_{ab} = \frac{1}{N}\eta_{abcd}E_I^c Q^d \,. \label{2foz}
\end{equation}
Finally, using an arbitrary timelike vector and 
expressions (\ref{pdfo}) we will find the remaining principal direction 
of the Papapetrou field.

We finish this case with the study of the following remaining subcase:

\vspace{2mm}

\noindent$\bullet\,${\em Particular subcase $\mb{E_R}=0$:}
The vanishing of $\mb{E_R}=0$ is equivalent to the relation
$\alpha\mb{\psi}+\beta\mb{\omega}=0$, which tells us that $\mb{\psi}$
and $\mb{\omega}$ are linearly dependent.  The difference that appears
between this subcase and the general case 
$E_R^aE^{}_{Ra}= 0$ is that now
we cannot determine explicitly any of the two principal directions,
since now we have $\mb{E_R}=\mb{P}=0$ [see (\ref{pdlp})].  
Therefore, like in the previous case, we have to find a timelike 
vector field and use the tensor ${\cal F}_{ab}$ (\ref{fgte}), 
which in this subcase has the following form
\[ {\cal F}_{ab}=-\frac{\alpha^2+\beta^2}{N}\left\{
\eta_{abcd}Q^c E_I^d+Q_aQ_b+E_{Ia}E_{Ib}-Ng_{ab}\right\} \,. \]
Finally, the principal null directions are given by expressions
(\ref{pdfo}). However, if we have a timelike vector orthogonal
to $\mb{E_I}$ and $\mb{Q}$, say $\mb{U}$ ($U^aE_{Ia}=U^aQ_a=0$), 
the principal directions can be found
through the following expressions
\[K_a = \eta_{abcd}U^bQ^c E_I^d-NU_a \,, \hspace{8mm}
  L_a = \eta_{abcd}U^bQ^c E_I^d+NU_a \,. \]

\subsection{Singular Papapetrou fields\label{snc1}}
As we have said before, this case is characterized by the vanishing
of the complex function $\alpha+i\beta$, which is equivalent to the
vanishing of the invariants of the Papapetrou field $x$ and $y$
(\ref{dexy}):
\begin{equation}
\psi^a\psi_a = \omega^a\omega_a \, , \hspace{15mm}
\psi^a\omega_a = 0 \, . \label{sico}
\end{equation}
Moreover, the exterior derivative of the Ernst potential
$\mb{\epsilon}$ (\ref{papk}) is a null 1-form,
$\epsilon^a\epsilon_a=0$, and the tensor ${\cal F}_{ab}$
becomes $-T_{ab}$, that is, it is symmetric, in agreement with the
fact that for singular 2-forms there is only one principal direction.  
It is obvious that in this case we cannot introduce the 1-form $\mb{E}$, 
but by virtue of (\ref{sico}), the exterior derivative of the Ernst
potential $\mb{\epsilon}$ is a complex 1-form satisfying similar
properties that $\mb{E}$, in the sense that the real and imaginary
parts of $\mb{\epsilon}$ are also orthogonal.

In order to determine the principal direction of a singular Papapetrou
field two subcases must be distinguished: (i) When $\mb{\psi}$ and
$\mb{\omega}$ are two spacelike 1-forms which are mutually orthogonal.
(ii) The rest of possibilities, which are characterized by the fact that
the 1-form $\mb{\epsilon}$, which is a null complex 1-form, is
proportional to a real null 1-form $\mb{k}$.

\subsubsection{Subcase (i):}

As we have said before, we consider $\mb{\psi}$ and
$\mb{\omega}$ to be spacelike and orthogonal. They cannot be timelike,
since (\ref{sico}) would imply that they should be null 1-forms, which
corresponds to the subcase (ii).  On the other hand, either $\mb{\xi}$ 
or the 1-form $\mb{\phi}$ must be timelike. Then, we can
determine the principal null direction by contracting the energy-momentum
tensor with one of them.  We find that this principal direction is
given by
\begin{equation}
\mb{k}= \psi^a\psi_a \mb{\xi} - N \mb{\phi} \,. \label{pdsc} 
\end{equation}

\subsubsection{Subcase (ii):}

This case, characterized by  
\begin{equation}
\mb{\epsilon} = \lambda \mb{k} \,,  \label{udpn}
\end{equation}
where $\lambda$ is a complex function, includes some particular cases
but in all of them $\mb{\psi}$ and $\mb{\omega}$ are null or
vanishing 1-forms. As we can see from expression (\ref{papf}), the null
vector $\mb{k}$ in (\ref{udpn}) determines the principal null
direction of the Papapetrou field. Finally, it is 
important to point out that in this case the principal null 
direction is orthogonal to the KVF $\mb{\xi}$, $k^a\xi_a=0$.

\section{Some properties of the principal directions of the Papapetrou 
fields\label{prop}}

We devote this section to study some properties and features
of the principal null directions of the Papapetrou field in vacuum
space-times.  More specifically, we study under which conditions these 
null principal directions are geodesic, and in that case we compute 
the optical scalars associated with them.  Here, we will also 
distinguish between the regular and singular cases.

\subsection{Regular Papapetrou fields}
First of all, we are going to study when a principal direction of 
the Papapetrou field is geodesic.  We can do that by using the
explicit expressions for the principal directions given in the previous
section, or their properties in the case $\mb{E_R}=0$, since in this 
case we do not know them explicitly. Then, the condition of geodesicity 
for a principal null direction is equivalent to say that the components 
of the vector $k^b k^a{}_{;b}$ on the orthogonal 2-planes must 
vanish.  Then, contracting this vector with the explicit expressions
for the generators of the orthogonal 2-planes (excepting in the
case $\mb{E_I}=0$, since in this case we do not have explicit 
expressions for them) and
making some straightforward calculations, we can establish the
conditions that a principal direction $\mb{k}$ must satisfy in
order to be geodesic.  The main point in these calculations is the
use of the following relation between $\mb{P}$, $\mb{Q}$ and $\mb{\xi}$
\begin{equation} 
\mb{Q}-\mb{P}=\mb{\xi} \,, \label{rqpx}
\end{equation}
which follows from (\ref{deeq}).  This expression allows us to 
transform terms with $\mb{Q}$ into terms with $\mb{P}$ [and hence, 
into terms with the principal directions through (\ref{pdlp})] 
and $\mb{\xi}$.  In this way, we can arrive at the following theorem:

\vspace{2mm}

\noindent {\bf Theorem 1:} Let $(V_4,g)$ be a vacuum space-time 
and let $\mb{\xi}$ be a KVF.  Then, a principal direction 
$\mb{k}$ of the Papapetrou field associated with $\mb{\xi}$ is 
geodesic if and only if the following condition holds
\[ \begin{array}{ll} 
k^ak^b E_{a;b} = 0 \hspace{5mm} & \mbox{when $\mb{E_I}\neq 0$} 
\,, \\ m^a (\alpha+i\beta)_{,a} = 0 & \mbox{when $\mb{E_I}=0$} \,.
\end{array} \]

\vspace{2mm}

Here, some remarks are in order.  First, $\mb{m}$ is an arbitrary
complex 1-form lying on the orthogonal 2-planes and such that 
$\mb{m}\wedge\mb{\bar{m}}\neq 0$.  We can find it by means of
the tensor ${\cal F}^\bot_{ab}$ (\ref{fort},\ref{tenf}).  Second, when 
$\mb{E_R}$ is a null 1-form we have seen that it is also one of the 
principal directions, and from expressions (\ref{ern1},\ref{ern2})
we deduce that $\mb{E_R}$ is in addition integrable, and therefore, 
it is geodesic.
Finally, it is important to note that in the case $\mb{E_I}\neq 0$,
the information on the geodesicity of a principal direction is 
encoded in a complex scalar depending only on the principal direction 
itself and the 1-form $\mb{E}$. In the case $\mb{E_I}=0$, it depends
on the gradient of $\alpha+i\beta$ on the orthogonal 2-planes.

\vspace{2mm}

When one of the principal directions is geodesic,
we can study its optical scalars: expansion, shear and rotation
(they are only well defined in the case of geodesic null vector
fields).  We begin with the shear of the geodesics tangent to
a given principal direction.  As is well known, if we choose
the tangent vector field $\mb{k}$ to be affinely parametrized,
then the shear scalar coincides with the quantity $\sigma\bar{\sigma}$, 
where $\sigma$ is one of the Newman-Penrose spin coefficients 
(see~\cite{NEPE,KSHM}).  But in general, we can take a 
Newman-Penrose basis associated with $\mb{k}$, namely 
\npb, and then to compute the
shear tensor ${\cal S}_{ab}$
\begin{equation}
{\cal S}_{ab}\equiv H_a{}^cH_b{}^d k_{(c;d)}-
\frac{1}{2}H_{ab}H^{cd}k_{c;d}\,,\hspace{6mm} 
H_{ab}=2m_{(a}\bar{m}_{b)}\,,\hspace{2mm} H^a{}_a=2\,,
\label{shte}
\end{equation}
where $H_{ab}$ is the orthogonal projector to the 2-planes
spanned by $\{\mb{k},\mb{\ell}\}$.  Then, taking into account
that the shear tensor ${\cal S}_{ab}$ is spacelike, the shear scalar 
${\cal S}^2\equiv \ts{1\over2}{\cal S}^{ab}{\cal S}_{ab}$ vanishes 
if and only if ${\cal S}_{ab}$ does, and it coincides with 
$\sigma\bar{\sigma}$ when $\mb{k}$ is chosen to be affinely 
parametrized. 

For our purposes it is better to compute the shear tensor 
(\ref{shte}).  In the case $\mb{E_I}\neq 0$ we can take $\mb{m}$ to 
be parallel to $\mb{E_I}+i\mb{Q}$ (with a proportional factor such 
that $m^a\bar{m}_a=1$), whereas in the second case it can be found 
through the tensor ${\cal F}^\bot_{ab}$ (\ref{fort},\ref{tenf}).  
Then, after some long but straightforward calculations and using 
here also the relation (\ref{rqpx}), we arrive to the following result
\[{\cal S}^2={\cal Z}\bar{{\cal Z}}\,,\hspace{5mm}\mbox{where}
\hspace{5mm} {\cal Z}= \left\{ \begin{array}{ll} 
\frac{1}{E_I^cE^{}_{Ic}}k^a E^b E_{[a;b]} \hspace{5mm} & \mbox{if 
$\mb{E_I}\neq 0$} \,, \\ [4mm]
\frac{k^cE_{Rc}}{N} m^a m^b E_{Ra;b} = 0 
\hspace{5mm} & \mbox{if $\mb{E_I}=0$} \,.
\end{array} \right. \]
Then, using this result we can enounce the next theorem:

\vspace{2mm}

\noindent {\bf Theorem 2:} Let $(V_4,g)$ be a vacuum space-time 
and let $\mb{\xi}$ be a KVF.  Then, a {\em geodesic} principal 
direction $\mb{k}$ of the Papapetrou field associated with $\mb{\xi}$ 
is shear-free if and only if the following condition holds 
\[ \begin{array}{ll} 
k^a E^b E_{[a;b]} = 0 \hspace{5mm} & \mbox{when $\mb{E_I}\neq 0$}\,,\\
m^a m^b E_{Ra;b} = 0 \hspace{5mm} & \mbox{when $\mb{E_I}=0$} \,.
\end{array} \]

\vspace{2mm}

Here it is important to point out that the condition for the case 
$\mb{E_I}\neq 0$ involves only 
antisymmetric covariant derivatives and therefore, this calculation
does not need the use of connection, we can make it by using only
partial differentiation. In fact, it is possible to show that 
this condition can be written as follows
\[ \left[(k^bE_b)E^a+Nk^a\right](\alpha+i\beta)_{,a}=0 \,\]
that is, in terms of derivatives of the eigenvalues of the Papapetrou
field.
Finally, taking into account the Goldberg-Sachs theorem~\cite{GOSA},
which states that a vacuum space-time is algebraically special if and
only if it contains a shearfree geodesic congruence, the next
corollary follows immediately

\vspace{2mm}

\noindent {\bf Corollary:} Let $(V_4,g)$ be a vacuum space-time 
and let $\mb{\xi}$ be a KVF.  If one of the principal directions 
of the Papapetrou field associated with $\mb{\xi}$, namely $\mb{k}$, 
satisfies the following conditions
\[  \begin{array}{ll}
k^a k^b E_{a;b} = 0 \hspace{3mm} \mbox{and} \hspace{3mm}
k^a E^b E_{[a;b]} = 0 \hspace{5mm} & \mbox{when $\mb{E_I}\neq 0$}\,,\\
m^a(\alpha+i\beta)_{,a} = 0 \hspace{3mm} \mbox{and} \hspace{3mm}
m^a m^b E_{Ra;b} = 0 \hspace{5mm} & \mbox{when $\mb{E_I}=0$}\,, 
\end{array} \]
the space-time is algebraically special and $\mb{k}$ is a 
principal direction of the Weyl tensor.

\vspace{4mm}

Now, let us study the expansion and rotation in the case of 
geodesic principal directions of the Papapetrou field. These
quantities are given by
\[ \vartheta \equiv \ts{1\over2}H^{ab}k_{a;b}\,, \hspace{10mm}
H_a{}^cH_b{}^d k_{[c;d]} \equiv -2i\varpi m_{[a}\bar{m}_{b]} \,,\]
where $\vartheta$ and $\varpi$ are the expansion and rotation scalars 
respectively.  In order to compute them it is more
appropriate to consider the complex scalar
\begin{equation} 
\rho \equiv -(\vartheta+i\varpi) = -m^a\bar{m}^bk_{a;b} \,,
\label{codi}
\end{equation}
usually called the {\em complex divergence} scalar, which is 
another spin coefficient in the Newman-Penrose formalism. 
The result of this calculation is 
\[ \rho = \left\{ \begin{array}{ll} 
\frac{1}{E_I^cE^{}_{Ic}}\left( k^aE^bE_{[a;b]}+k^aE_R{}^bE_{a;b}
\right) & \mbox{if $\mb{E_I}\neq 0$}\,, \\ [4mm]
-\frac{1}{N}\left[(k^cE_{Rc})m^a\bar{m}^bE_{R(a;b)} + 
(k^c\xi_c)W_{ab}m^a\bar{m}^b\right] & \mbox{if $\mb{E_I}=0$}\,. 
\end{array} \right. \]
In the case $\mb{E_I}=0$ we can identify immediately the expansion
and shear scalars, they are
\begin{equation} 
\vartheta=\frac{k^cE_{Rc}}{N}m^a\bar{m}^bE_{R(a;b)}\,, \hspace{8mm}
\varpi=-i\frac{k^c\xi_c}{N}W_{ab}m^a\bar{m}^b\,. \label{exro}
\end{equation}
Here, it is interesting to note the direct relationship between the 
rotation $W_{ab}$ of $\mb{\xi}$ and the rotation $\varpi$ of the 
principal direction.
With regard to the case $\mb{E_I}\neq 0$, it is remarkable 
that when the geodesic principal direction $\mb{k}$ is also shear-free, 
the complex divergence is simply
\[ \rho = \frac{1}{E_I^cE^{}_{Ic}} k^aE_R{}^bE_{a;b} \,, \]
and hence, the expansion and rotation scalars are given by
\[ \vartheta = \frac{1}{E_I^cE^{}_{Ic}} k^aE_R{}^bE_{Ra;b} \,, 
\hspace{6mm} \varpi = \frac{1}{E_I^cE^{}_{Ic}} k^aE_R{}^bE_{Ia;b} \,. \]
Finally, it is important to remark the role that the 1-form $\mb{E}$
plays also in the differential properties of the principal null
directions of a regular Papapetrou field, since as we can see in all
the expressions of this section, the optical scalars as well as 
the geodesicity condition are given only in terms of scalars formed
from the principal direction $\mb{k}$ and the covariant derivative
of $\mb{E}$.

\subsection{Singular Papapetrou fields}

The general situation in the singular case can be
summarized as follows: we have a vacuum space-time containing a 
singular 2-form (the Papapetrou field), which is a solution of 
the Maxwell equations (\ref{maxe}). Therefore, the Mariot-Robinson 
theorem~\cite{MARI,ROBI} (see also~\cite{KSHM}) tells us that this is
equivalent to say that there
is a geodesic and shear-free congruence of null curves.  In addition,
the principal null direction of the Papapetrou field will be tangent
to these curves.  Finally, the Mariot-Robinson theorem tell us that the
space-time is algebraically special and then, the following theorem 
follows

\vspace{2mm}

\noindent {\bf Theorem 3:} Let $(V_4,g)$ be a vacuum space-time and let 
$\mb{\xi}$ be a KVF.  If the gradient of the Ernst potential 
$\mb{\epsilon}$ is a complex null 1-form, then the space-time will 
be algebraically special and the principal null direction of the
 
singular Papapetrou field associated with $\mb{\xi}$ will be geodesic 
and shear-free.

\vspace{2mm}

Taking into account that we are dealing with vacuum space-times, this
theorem tells us that if the gradient of any KVF in a vacuum space-time 
is a null 1-form,  then the space-time must be algebraically special 
and a multiple direction of the Weyl tensor
coincides with the null principal direction of the corresponding
Papapetrou field.

On the other hand, for the subcase (i), the complex divergence
$\rho$ (\ref{codi}) is given by the following expression
\[ \rho = \frac{1}{2\psi^c\psi_c}k^a\bar{\epsilon}^b\epsilon_{b;a}
\,. \]
In the subcase (ii), using the equation (\ref{eqep}), we can see
that $\rho=0$.

\section{Some examples\label{exam}}

In this section we apply our previous study to some examples.  
In particular, we study the Papapetrou field associated with KVFs in 
the following vacuum space-times: (i) The Kerr metric. (ii) The Kasner 
metrics. (iii) A subclass of the plane-fronted gravitational waves.   
In these examples we compute the principal direction(s) of the 
Papapetrou field and also discuss some of their algebraic 
and differential properties. 

\vspace{4mm}

In the first example we consider the well-known Kerr 
metrics~\cite{KERR}.  
They constitute a family of stationary axisymmetric vacuum solutions 
of Einstein's field equations.  In Boyer-Lindquist (BL) coordinates 
$\{t,r,\theta,\varphi\}$ (see~\cite{BOLI,KSHM}), the line-element 
takes the form
\begin{equation}
\fl ds^2 =  -\frac{\Delta}{\rho^2}(dt-a\sin^2\theta d\varphi)^2+
\frac{\sin^2\theta}{\rho^2}\left[(r^2+a^2)d\varphi-adt\right]^2+
\frac{\rho^2}{\Delta} dr^2 + \rho^2 d\theta^2 \, , \label{kerr}
\end{equation}
where $\Delta$ and $\rho$ are given by $\Delta = r^2-2Mr+a^2$
and $\rho^2 = r^2+ a^2 \cos^2 \theta$ respectively. Here, we
must say that in the case $a^2<M^2$, the BL coordinates are 
valid in the asymptotically flat regions ($r_+<r<\infty$) or type I 
regions, in the type II regions ($r_-<r<r_+$), which contain closed 
trapped surfaces, and in the asymptotically flat regions containing 
the ring singularity $\rho^2=0$ ($-\infty<r<r_-$) or type III regions
($r_\pm = M\pm(M^2-a^2)^{1/2}$; see \cite{HAEL} for more details).
The only singularity of this metric is the ring singularity $\rho^2=0$.
The event horizons ($\Delta=0$\;$\Leftrightarrow$\;$r=r_\pm$) and
the axis of symmetry ($\theta=0$) are just coordinate singularities,
as is well-known, we can avoid them by chosing other coordinate
systems.

This metric has an Abelian two-parameter group of isometries.  We
can choose two independent KVFs in the following way
\begin{equation} 
\vec{\mb{\xi}}_{\mb{\scriptstyle t}} = 
\frac{\mb{\partial}}{\mb{\partial t}}\,,\hspace{8mm} 
\mb{\xi_t} = -\left(1-\frac{2Mr}{\rho^2}\right)\mb{dt}-
\frac{2Mar\sin^2\theta}{\rho^2}\mb{d\varphi} \,, \label{kite}
\end{equation}
\begin{equation}
\vec{\mb{\xi}}_{\mb{\scriptstyle \varphi}} = 
\frac{\mb{\partial}}{\mb{\partial \varphi}}\,,\hspace{8mm}
\mb{\xi_\varphi}= -\frac{2Mar\sin^2\theta}{\rho^2}\mb{dt}+
\frac{\Sigma^2\sin^2\theta}{\rho^2}\mb{d\varphi} \,, 
\label{kias}
\end{equation}
where $\Sigma^2\equiv (r^2+a^2)^2-a^2 \Delta \sin^2\theta$. 
Now, we will study the Papapetrou field associated with the 
timelike KVF, $\mb{\xi_t}$,
\begin{eqnarray}
\mb{F} = \frac{2M}{\rho^4}(r^2-a^2\cos^2\theta)\left(\mb{dt}
-a\sin^2\theta\mb{d\varphi}\right)\wedge\mb{dr} \nonumber \\
+ \frac{2Ma r\sin(2\theta)}{\rho^4}\mb{d\theta}\wedge\left(
a\mb{dt}-(r^2+a^2)\mb{d\varphi}\right) \,. \label{pfkt}
\end{eqnarray}
It is important to point out (see~\cite{WALD}) that this Papapetrou 
field is proportional to the electromagnetic field of the Kerr-Newman 
metric (see also~\cite{MTWB} for a detailed study of this 
electromagnetic field). As we can see from (\ref{kite}), the norm 
of $\mb{\xi_t}$ is given by
\begin{equation} 
N = -\left(1-\frac{2Mr}{\rho^2}\right) = 
-\frac{\Delta-a^2\sin^2 \theta}{\rho^2} \,. \label{norm} 
\end{equation}
As is clear, it vanishes on the surfaces of infinite redshift 
($\Delta=a^2\sin^2\theta$), and in principle we cannot apply our 
procedure there.  However, this situation is different from that 
of~\ref{appa}, and later we will see that it is possible to use 
the general procedure of Sec.~\ref{dosf}.

In the regions where the BL coordinates are valid, we find the 
following expressions for $\mb{\psi}$ and $\mb{\omega}$
\[ \mb{\psi} = \frac{2M}{\rho^4}\left\{(r^2-a^2\cos^2\theta)\mb{dr}-
a^2\sin(2\theta)r\mb{d\theta}\right\} \, ,  \]
\[ \mb{\omega} = \frac{2Ma\sin\theta}{\rho^4}\left\{\cot\theta\mb{dr}+
(r^2-a^2\cos^2\theta)\mb{d\theta}\right\} \, ,  \]
and then, we can write the Ernst potential as follows
\[ {\cal E} = -\frac{2M}{\rho^2}(r+ia\cos\theta) = -
\frac{2M}{r-ia\cos\theta} \,. \]
The eigenvalues of the Papapetrou field (\ref{pfkt}) are given by
\begin{equation} 
\alpha+i\beta = s\frac{2M}{(r-ia\cos\theta)^2} = 
s\frac{{\cal E}^2}{2M} \,, \label{amib}
\end{equation}
and therefore, it is regular everywhere excepting in the ring 
singularity $\rho^2=0$, where it diverges. On the other hand, it is
remarkable that $\alpha+i\beta$ is an analytic function of 
the Ernst potential.  As we have pointed out before, this can only 
happens when $\mb{dE} = 0$ [see equation (\ref{defe})], like in 
this example, where the 1-form $\mb{E}$ is given by
\[ \mb{E} = s (\mb{dr}+ia\sin\theta\mb{d\theta}) \hspace{6mm}
\Longrightarrow \hspace{6mm} \mb{dE} = 0 \,. \]
Moreover,
\[ \fl \mb{E_R} = s\mb{dr} \,, \hspace{4mm} E_R^aE^{}_{Ra}=
\frac{\Delta}{\rho^2} \,, \hspace{8mm} \mb{E_I}=
sa\sin\theta\mb{d\theta} \,, \hspace{4mm} E_I^aE^{}_{Ia}=
\frac{a^2\sin^2\theta}{\rho^2}\,,\]
and hence, excepting in $r=r_\pm$ ($\Delta=0$) and in the axis 
of symmetry ($\theta=0$), 
the Papapetrou field corresponds to the generic case (Subsection 
\ref{sbgc}).  In order to compute the principal directions
of the Papapetrou field (\ref{pfkt}) in the generic case, we need 
to compute the 1-form $\mb{D}$
\[ \mb{D} = -\frac{a\Delta\sin^2\theta}{\rho^2}\mb{d\varphi} \,, 
\hspace{6mm} D^a D_a =\frac{a^2\Delta\sin^2\theta(\Delta-a^2 
\sin^2\theta)}{\rho^6} \,. \]
Then, $\mb{P}$ and $\mb{Q}$ are
\[ \mb{P} = \frac{\Delta}{\rho^2}(\mb{dt}-a\sin^2\theta\mb{d\varphi})\,
\hspace{1cm} \mb{Q} = \frac{a\sin^2\theta}{\rho^2}(a\mb{dt}-(r^2+a^2)
\mb{d\varphi}) \,, \]
and finally, the principal null directions of the Papapetrou field are
[see (\ref{pdlp})]
\begin{equation}
\fl \mb{K} = - \frac{\Delta}{\rho^2}\mb{dt}+s\mb{dr}+  
\frac{a\sin^2\theta \Delta}{\rho^2} \mb{d\varphi} \,, \hspace{10mm}
\mb{L}= -\frac{\Delta}{\rho^2}\mb{dt}-s\mb{dr}+  
\frac{a\sin^2\theta \Delta}{\rho^2} \mb{d\varphi} \,, \label{dirk}
\end{equation} 
which are just the multiple principal directions\footnote{Note that 
the arbitrary sign $s$ only serves to change from $\mb{K}$ to $\mb{L}$ 
and the converse.} of the Kerr space-time
(see for instance~\cite{CHAN}), whose Petrov type is D.  We can
check, using Theorems 1 and 2, that $\mb{K}$ and $\mb{L}$ are
indeed geodesic and shearfree.

On the other hand, in~\ref{appb} we have given the connection with
the {\em eigenray} formalism of Perj\'es, showing that the existence
of a geodesic (and shear-free) principal direction of the Papapetrou
field is equivalent to the existence of a geodesic (and shear-free)
eigenray.   This result, together with the fact that the Kerr 
space-time is the only asymptotically-flat stationary vacuum solution 
with geodesic and shear-free eigenrays~\cite{PER2}, shows that
the Kerr metric is the only one with such characteristics in which 
the alignment of the principal directions of the Papapetrou 
and gravitational fields takes place, which has been shown 
recently by Mars~\cite{MARS}.  In this paper, the characterization
is given in the whole Kerr manifold, whereas the result of 
Perj\'es~\cite{PER2} was found only in the region where the KVF
is timelike.  However, taking into account that the eigenray
formalism can be used also in the case of spacelike KVFs, Perjes' 
result could be extended to the whole manifold.

Now, in order to show the capabilities of our formalism, we will extend 
the analysis to the symmetry axis, the even horizons, 
and the surfaces of infinite redshift. On the axis of symmetry (where 
$\mb{E_I}=0$ and $E_R^aE^{}_{Ra}\neq 0$) we can find the principal 
directions by using expressions (\ref{pdiv}).  However, the BL 
coordinates are not good on the axis.  In order to solve this problem 
we can make the calculation in Kerr-Schild coordinates 
(see~\cite{KSHM,HAEL}).  It is easy to see that the principal directions 
obtained can be expressed in terms of the coordinates $\{t,r\}$, they
are given by
\[ \mb{K} = -\left(1-\frac{2Mr}{r^2+a^2}\right)\mb{dt}+
s\mb{dr}\,, \hspace{8mm} \mb{L}= -\left(1-\frac{2Mr}
{r^2+a^2}\right)\mb{dt}-s\mb{dr}  \,. \]

The surfaces of infinite redshift ($\Delta=a^2\sin^2\theta$)
constitute a controversial point in this example, because there the 
norm of $\mb{\xi_t}$ vanishes [see (\ref{norm})]. However, 
$\alpha+i\beta\neq 0$ (\ref{amib}) is different from zero on these 
surfaces, which tells us that we are not in the case of~\ref{appa}.
Then, we can follow the general procedure explained in 
Sec.~\ref{dosf}.  In this case, the Papapetrou field belongs to
the generic case, and we can see that the principal directions 
are given by (\ref{dirk}).

In the case of the event horizons, we are going to consider the 
maximal extension of the Kerr metric.  As is well-known, it can be 
constructed by using advanced and retarded Kerr coordinates 
$\{V_\pm,r,\theta,\varphi_\pm\}$ ($-\infty<
V_\pm<\infty$, $-\infty<r<\infty$, $0\leq \theta< \pi$,
$0\leq \varphi_\pm<2\pi$)
\[ dV_\pm = dt \pm (r^2+a^2)\Delta^{-1}dr\,, \hspace{6mm}
d\varphi_\pm = d\varphi \pm a\Delta^{-1}dr \,. \]
Now, the line element is analytic (apart from regions I, II, and
III) also in $r=r_\pm$.  Therefore, we can already compute
the principal directions on $r_\pm$,  where we have 
$E_R^aE^{}_{Ra}=0$ and $\mb{E_R}\neq 0$.  Taking into 
account the procedure explained in Sec.~\ref{sepf}, one
of the principal directions is given by $\mb{E_R}=\mb{dr}$,
and the other one is obtained by contracting (\ref{fgte}) with
any timelike vector field.  In our case, we find that 
on the event horizons $\mb{{\cal F}}$ is given by
\[ \mb{{\cal F}} =  2(\mb{dV_\pm}-a\sin^2\theta\mb{d\varphi_\pm})
\otimes \mb{dr} \,,\]
and therefore, the other principal direction is given by
$\mb{dV_\pm}-a\sin^2\theta\mb{d\varphi_\pm}$ (the vector field
associated with this 1-form is $\mb{\partial/\partial r}$,
which is tangent to the curves $V_\pm=constant$).

To sum up, the maximal extension of the Kerr metric (see 
\cite{HAEL} for details), in the case $a^2<M^2$, is made up of an
infinite chain of asymptotically-flat type I regions,
connected to type II regions (which contain trapped
surfaces) and asymptotically-flat type III regions
(which contain the ring singularity) by means of the null 
hypersurfaces $r=r_\pm$.  We have just seen that 
we can compute the principal directions of the Papapetrou field 
(\ref{pfkt}) for the whole Kerr manifold, and in
addition, that they coincide with the repeated principal
directions of the Weyl tensor.  For the sake of 
completeness, we want to remark that these results can
be trivially extended to the cases $a^2=M^2$ (where
$r_+=r_-$ and there is no region II) and $a^2>M^2$
(where $\Delta>0$ everywhere).

We finish this example with a comment on the other KVF, 
$\mb{\xi_\varphi}$ (\ref{kias}). Both KVFs are privileged since
$\mb{\xi_t}$ is the only one (up to a multiplicative constant)
which is timelike at arbitrary large positive and
negative values of $r$, and $\mb{\xi_\varphi}$ is the only one
which vanishes on the axis of symmetry ($\theta=0$) and
satisfies the regularity condition there (see~\cite{KSHM}).
However, we can check that the principal directions of the
Papapetrou field associated with $\mb{\xi_\varphi}$ do not
coincide with the principal directions of the Weyl tensor.
In this sense, our study privileges the KVF $\mb{\xi_t}$
over the KVF $\mb{\xi_\varphi}$.

\vspace{4mm}

The next example will be an algebraically general vacuum space-time.  
One of the most simple families of this type is the following class
of Kasner metrics~\cite{KASN} (see also~\cite{KSHM})
\[ ds^2 = -dt^2+t^{2p_1}dx^2+t^{2p_2}dy^2+t^{2p_3}dz^2 \, , \]
where $p_1$, $p_2$ and $p_3$ are constants satisfying the following
relationships
\[ p_1+p_2+p_3=1 \,, \hspace{1cm} p^2_1+p^2_2+p^2_3 = 1 \,. \]
This metric has an Abelian $G_3$ group of isometries acting on the
spacelike hypersurfaces $\{t=\mbox{constant}\}$.  We can choose one
of these KVFs, for instance 
\[ \vec{\mb{\xi}}_{\mb{\scriptstyle x}} \equiv 
\frac{\mb{\partial}}{\mb{\partial x}}\,, \hspace{5mm} 
\mb{\xi_x} = t^{2p_1} \mb{dx} \hspace{5mm} \Longrightarrow 
\hspace{5mm} \mb{F} = 2p_1 t^{2p_1}\mb{dt}\wedge\mb{dx} \,.\]
Then, 
\[ N = t^{2p_1}\,,\hspace{4mm} \mb{\psi}=-2p_1 t^{2p_1-1}\mb{dt}\,,
\hspace{1cm}\mb{\omega}=0\,,\]
which means that the Ernst potential associated with $\mb{\xi_x}$ 
is simply ${\cal E} = -t^{2p_1}$.
Now, if we compute the eigenvalues of the Papapetrou field
\[ \alpha = 2sp_1 t^{p_1-1} \, , \hspace{1cm} \beta =0 \,, \]
we realize that it is a regular electromagnetic field, but it does not
belong to the generic class, since 
\[ \mb{E} = -st^{p_1}\mb{dt} \hspace{5mm} \Longleftrightarrow 
\hspace{8mm} \mb{E_R}=-st^{p_1}\mb{dt}\,,\hspace{4mm}\mb{E_I}=0\,.\]
That is to say, we are in the particular case $\mb{E_I}=0$, and 
therefore the principal directions of the Papapetrou field are given by 
(\ref{pdiv})
\begin{equation} 
\vec{\mb{K}} = s t^{p_1}\frac{\mb{\partial}}{\mb{\partial t}} +
\frac{\mb{\partial}}{\mb{\partial x}} \,, \hspace{6mm}
\vec{\mb{L}} = -s t^{p_1}\frac{\mb{\partial}}{\mb{\partial t}} +
\frac{\mb{\partial}}{\mb{\partial x}} \,. \label{pdkm}
\end{equation}
The complex vector $\mb{m}$ which determines
the orthogonal 2-planes to the principal directions, which is  
obtained from (\ref{fort}) or (\ref{tenf}), can be taken as follows
\[ \vec{\mb{m}} = \frac{1}{\sqrt{2}}\left(t^{-p_2}\frac{\mb{\partial}}
{\mb{\partial y}}+it^{-p_3}\frac{\mb{\partial}}{\mb{\partial z}} 
\right) \,. \]
Now, we can check that the null principal directions (\ref{pdkm}) of the
Papapetrou field satisfy the condition of the Theorem 1, and therefore 
they are geodesic.  In the general case, they do not coincide with the 
Weyl principal directions.   Moreover, they satisfy the condition of the 
Theorem 2 only when $p_2 = p_3$, in which case they are also shearfree 
and hence, via the Goldberg-Sachs theorem~\cite{GOSA}, the space-time is 
algebraically special (specifically, Petrov type D), being (\ref{pdkm}) 
the repeated null principal directions.  Finally, using expressions 
(\ref{exro}), we find that the expansion of $\mb{k}$ and $\mb{\ell}$ 
is given by
\[ \vartheta_\pm = \pm\frac{1}{2}s(p_1-1)t^{p_1-3} \,, \]
(the signs $+$ and $-$ are for $\mb{k}$ and $\mb{\ell}$ respectively),
and the rotation is zero since the rotation of $\mb{\xi_x}$ also 
vanishes [see (\ref{exro})].

\vspace{4mm}

Another interesting example is a subfamily of the plane-fronted 
gravitational waves, {\em pp waves} (see for instance~\cite{KSHM}).
The line-element of these space-times is 
\[ ds^2 = -2du dv + 2d\zeta d\bar{\zeta} - 2H du^2 \,, \]
where 
\[ H = f(\zeta)+\bar{f}(\bar{\zeta}) \,,  \]
that is, this is the subclass of the vacuum plane waves with the 
additional KVF
\[ \vec{\mb{\xi}}_{\scriptstyle u}= \frac{\mb{\partial}}{\mb{\partial u}}
\,, \hspace{5mm} \mb{\xi_u} = -\mb{dv}-2 H \mb{du}\,, \hspace{1cm} 
N = -2H = -2\left(f(\zeta)+\bar{f}(\bar{\zeta}) \right) \,.\] 
(These metrics have always
$\mb{\partial/\partial v}$ as a KVF but it is a null KVF; see 
Appendix A).   The Papapetrou field associated with $\mb{\xi_u}$
is given by 
\begin{equation} 
\mb{F} = 2\mb{du}\wedge\left(H_{\!,\zeta}\mb{d\zeta}+
H_{\!,\bar{\zeta}}\mb{d\bar{\zeta}}\right) \,, \label{pwpf}
\end{equation}
Then, the 1-forms $\mb{\psi}$ and $\mb{\omega}$ are given by
\begin{equation} 
\mb{\psi} = 2\left(H_{\!,\zeta}\mb{d\zeta}+H_{\!,\bar{\zeta}}
\mb{d\bar{\zeta}}\right)\,, \hspace{5mm} \mb{\omega} =
-2i\left(H_{\!,\zeta}\mb{d\zeta}- H_{\!,\bar{\zeta}}
\mb{d\bar{\zeta}}\right)\,. \label{pw1f} 
\end{equation}
From these expressions we can write the Ernst potential associated with
$\mb{\xi_u}$ as follows
\[ {\cal E} = 4 f(\zeta)  \,. \]
Furthermore, from (\ref{pw1f}) we deduce that 
\[ \psi^a\psi_a = \omega^a\omega_a = 8 H_{\!,\zeta}H_{\!,\bar{\zeta}}
\geq 0\,, \hspace{10mm} \psi^a\omega_a = 0 \,. \]
Then, the Papapetrou field (\ref{pwpf}) is singular ($\alpha+i\beta=0$), 
and it corresponds to the subcase (i) of the subsection~\ref{snc1}. 
Using (\ref{pdsc}), the principal null direction is given by 
\begin{equation} 
\vec{\mb{k}} = 16 H H_{\!,\zeta}H_{\!,\bar{\zeta}} 
\frac{\mb{\partial}}{\mb{\partial v}} \,, \label{typn}
\end{equation}
which is the propagating direction of the plane wave, orthogonal to
the wave fronts.   Moreover, it is proportional to the null KVF 
$\mb{\partial/\partial v}$ and hence, taking into account the Theorem 
3, $\mb{k}$ is geodesic and shear-free.  From the Goldberg-Sachs 
theorem~\cite{GOSA}, we deduce that it is also a principal null 
direction of the space-time.  In this case the metric is of the Petrov 
type N, with only one multiple principal direction, which is given by 
(\ref{typn}).  Finally, the complex divergence $\rho$ vanishes since 
$\mb{\partial/\partial v}$ is a constant vector field.

\section{Remarks and conclusions\label{reco}}

In this paper we have developed a formalism for the study of
the Papapetrou fields in vacuum space-times.  In the first part
of this paper, we have
determined the null principal direction(s).  In the case of
regular Papapetrou fields, this is very useful in order to
solve Maxwell's equations in curved space-times, since when we 
write them in a Newman-Penrose basis adapted to these principal 
directions the conditional differential system for them is 
second-order, contrary to what happens when we write them in 
an arbitrary Newman-Penrose basis, in which case it is a 
third-order differential system.   
In the second part of this work, we have studied the main 
differential properties of the null principal direction(s) of
the Papapetrou field.  Specifically, we have found the condition
that a principal direction must satisfy in order to be geodesic.
Moreover, for geodesic principal directions we have found
explicit expressions for the optical scalars.  This
study allows us to study when a principal direction of the
Papapetrou field is also a principal direction of an algebraically
special vacuum space-time.  
Furthermore, taking into account the simplicity of the expressions for
the differential properties, and the fact that they only depend on
the derivatives of the Ernst potential and the principal direction, they 
provide a interesting way of introducing new Ans\"atze for the search 
of exact solutions. The examples studied here can serve as a guide.  
As we have seen, the principal directions of the Kerr space-time 
are aligned with those of Papapetrou field associated with the
timelike KVF (see also~\cite{FASO,MARS}), which gives another 
characterization of this metric~\cite{MARS}, and the other
two cases (Kasner and {\em pp waves}) are also examples of how
the algebraic structure of the Papapetrou field can be adapted 
(for some KVFs) to the algebraic structure of the gravitational 
field. In this sense, the expressions for the expansion and rotation
scalars of the principal directions are also useful.

Finally, this study can be extended in several ways.  One of them is 
to study the relationship between the principal direction(s) of the 
Papapetrou fields and the possible algebraic types of the space-time.  
Here, there are several cases 
depending on the multiplicity and degree of alignment of the
principal direction(s) of the Papapetrou field~\cite{FSP2}.
On the other hand, taking into account that most of the properties
studied can be expressed in terms of objects related directly
to the Ernst potential, it seems reasonable to extend this study
to other space-times in which the Ernst potential can be defined,
for instance to Einstein-Maxwell space-times possessing a Killing
vector field.

\ack 
F.F. wishes to thank the D.G.R. of the Generalitat de
Catalunya (grant 1998GSR00015), and the Spanish Ministry 
of Education (contract PB96-0384). 
C.F.S. gratefully acknowledges financial support 
in the form of a fellowship from the Alexander von Humboldt 
Foundation.

\appendix 
 
\section{The case of a null Killing vector field\label{appa}}  

In this Appendix we sum up briefly what happens when
we consider a null KVF.  In order to study this case,
it is enough to remember that all the vacuum space-times admitting a 
null Killing vector field are known, they can be found in~\cite{KSHM} 
(section 21.4; see also Ref.~\cite{PER1}).  The Killing equations 
(\ref{kill}) tell us that the KVF is a geodesic, shear-free, and 
expansion-free null vector field, and the Einstein field equations 
imply the vanishing of the twist $\mb{\xi}\wedge\mb{d\xi} = \mb{0}$. 
As it was shown by Dautcourt~\cite{DAUT}, these space-times can be
divided into two different classes: 

-Class I is the family of the {\em pp} waves (see~\cite{KSHM}), 
in which the KVF is a constant null vector field 
($\xi_{a;b}=0$), and therefore the associated 
Papapetrou field is identically zero.  

-Class II is determined by the following line element 
(see~\cite{KSHM})
\[ ds^2 = -2 x du (dv+Mdu)+x^{-1/2}(dx^2+dy^2) \,, \]
where the function $M=M(u,x,y)$ satisfies the following partial
differential equation:
\[ (xM_{,x})_{,x}+xM_{,yy}=0\,.\]
In this case, the null KVF is 
\begin{equation}
\vec{\mb{\xi}}=\frac{\mb{\partial}}{\mb{\partial v}}\,,\hspace{1cm} 
\mb{\xi}=-x\mb{du}\,. \label{nknk} 
\end{equation}
Then, the associated Papapetrou field is simply 
$\mb{F}=\mb{du}\wedge\mb{dx}$. As is clear, it is singular and 
the corresponding principal direction is given by 
the null KVF (\ref{nknk}). Therefore, the principal null 
direction is geodesic and shear-free, and if we choose the KVF
(\ref{nknk}) to be the tangent vector field, the integral curves are 
affinely parametrized, being $v$ an affine parameter.  

\section{Connection with the {\em eigenray} formalism\label{appb}}

In reference~\cite{PER2} Perj\'es developed a spinor calculus
for stationary space-times from which, a triad formalism similar
to the Newman-Penrose formalism~\cite{NEPE} was put forward.  
An applications of this technique is the search of exact solutions 
of Einstein's equations.  To that end, Perj\'es introduced the 
concept of {\em eigenray}. Here, we show the connection between 
this concept and our development.

Let us consider a non-null KVF $\mb{\xi}$ and a null vector field 
$\mb{k}$ normalized by $k^a\xi_a=1$. These two objects determine a 
unit vector field $\mb{n}$ by
\begin{equation} 
\mb{n} = \sqrt{|N|}\,(\mb{k}-N^{-1}\mb{\xi}) \,, 
\hspace{8mm} n^an_a=-\mbox{sgn}(N) \,, \label{eray} 
\end{equation}
where $|x|$ and $\mbox{sgn}(x)$ denote the absolute value and the
sign of $x$ respectively.  Now, we can show~\cite{PER2} that
$\mb{n}$ is tangent to geodesics in the 3-space of the Killing 
orbits, which are called {\em eigenrays}, when it satisfies the 
following algebraic condition
\begin{equation} 
p^{ab}\psi_b+\varepsilon^a{}_{bc}n^b\omega^c=0\,,\hspace{10mm}
\varepsilon_{abc}\equiv(|N|)^{-1/2}\eta_{dabc}\xi^d\,,\label{eige}
\end{equation}
where $ p_{ab}\equiv g_{ab}+\mbox{sgn}(N)n_an_b$ is the
orthogonal projector to $\mb{n}$.  It can be also proven that 
condition~(\ref{eige}) is equivalent to say that $\mb{k}$ is geodesic
(see \cite{PER2}), therefore $\mb{n}$ is geodesic if and only if
$\mb{k}$ is geodesic. Moreover, Perj\'es showed that a geodesic eigenray
is also shear-free if and only if it is the null vector field 
associated through (\ref{eray}).

Now, let us consider the Papapetrou field associated with $\mb{\xi}$.  
It is possible to show that the null
vector field $\mb{k}$ in (\ref{eray}) is a principal direction of 
the Papapetrou field if and only if condition (\ref{eige})
holds. Therefore, this result shows that the existence of
a geodesic eigenray is equivalent to the existence of a geodesic
principal direction of the Papapetrou field (and the same for
the shear-free case).


\end{document}